\documentclass[%
superscriptaddress,
twocolumn,
10pt,
 amsmath,amssymb,
 aps,
 pra,
floatfix]{revtex4-1}

\usepackage{nicefrac}
\usepackage{graphicx}
\usepackage{dcolumn}
\usepackage{multirow}

\usepackage{color}

\newcolumntype{d}[1]{D{.}{.}{#1}}

\usepackage{pdfpages}
\setboolean{@twoside}{false}
\makeatletter
\AtBeginDocument{\let\LS@rot\@undefined}
\makeatother

\usepackage{hyperref}
\begin{document}
\title{Educational commitment and social networking:The power of informal networks}

\author{Justyna P. Zwolak}
\email{j.p.zwolak@gmail.com}
\affiliation{Department of Teaching and Learning, Florida International University, Miami, FL 33199, USA}
\affiliation{Joint Center for Quantum Information and Computer Science, UMD, College Park, MD 20742, USA}

\author{Michael Zwolak} 
\affiliation{Center for Nanoscale Science and Technology, National Institute of Standards and Technology, Gaithersburg, MD 20899}

\author{Eric Brewe}
\affiliation{Drexel University, Physics Department, Philadelphia, PA 19104, USA}
\affiliation{Drexel University, School of Education, Philadelphia, PA 19104, USA}

\date{\today}
\begin{abstract}
The lack of an engaging pedagogy and the highly competitive atmosphere in introductory science courses tend to discourage students from pursuing science, technology, engineering, and mathematics (STEM) majors. Once in a STEM field, academic and social integration has been long thought to be important for students' persistence. Yet, it is rarely investigated. In particular, the relative impact of in-class and out-of-class interactions remains an open issue. Here, we demonstrate that, surprisingly, for students whose grades fall in the ``middle of the pack,'' the out-of-class network is the most significant predictor of persistence. To do so, we use logistic regression combined with Akaike's information criterion to assess in- and out-of-class networks, grades, and other factors. For students with grades at the very top (and bottom), final grade, unsurprisingly, is the best predictor of persistence---these students are likely already committed (or simply restricted from continuing) so they persist (or drop out). For intermediate grades, though, only out-of-class closeness---a measure of one's immersion in the network---helps predict persistence. This does not negate the need for in-class ties. However, it suggests that, in this cohort, only students that get past the convenient in-class interactions and start forming strong bonds outside of class are or become committed to their studies. Since many students are lost through attrition, our results suggest practical routes for increasing students' persistence in STEM majors.
\end{abstract}
\maketitle

\section{Introduction}\label{sec:intro}
From industry to government to academia, attracting and retaining science, technology, engineering, and mathematics (STEM) majors is recognized as a key element of the 21$^{\text{st}}$ century knowledge economy~\cite{Adkins12-ANS,PCAST12-ETE,NSF96-STF}. One approach to attracting and retaining students is to improve the overall educational experience, which requires understanding the immersion of students into the academic and social system of an institution. For instance, it has been noted that insufficient interactions with others, as well as a lack of compatibility with the social values of the institution, lead to a low commitment to the university~\cite{Tinto75-DHE}. Ultimately, this affects one's decision about whether to drop out. For students who just started their education and have not yet formed connections in the community, particularly those who commute to college, the classroom might be the only place where connecting with others happens. However, while both social and academic involvement seem to be essential for persistence, research on the effect of students' networks within the university is rare~\cite{Thomas00-TTB,Dawson08-RSC,Zwolak17-IIP}.

The importance of the classroom experience in introductory courses as a means for improving students' persistence should not be underestimated. Nearly half of first-time students who leave a university by the end of the freshman year never come back to college~\cite{Swail04-ASR}. In a recent study, Zwolak {\it et al.} investigate academic and social experiences of students in an introductory physics classroom via social network analysis (SNA)~\cite{Zwolak17-IIP}. Their analysis of networks of self-reported, in-class interactions reveals that students with a higher number of initiated or received interactions (as measured by directed degrees) and higher overall embeddedness within the network (as measured by closeness) at the end of the first semester were more likely to persist, i.e., to enroll in a second course in the sequence. Out of all the examined indices, closeness was the most correlated, giving up to a 75\,\% chance of correctly predicting persistence. This agrees with earlier work that found closeness to be positively associated with students' perceptions of team effectiveness and performance, as well as on their attitudes and grades~\cite{Baldwin97-NES}.

In a seminal paper, Tinto noted that ``the manner in which social and academic involvements (integration) shape learning and persistence will vary~\dots\,for different students inside and outside the classroom.~\dots\,[N]etwork analysis and/or social mapping of student interaction patterns~\dots\,will shed important light on how interactions across the academic and social geography of a campus shape the educational opportunity structure of campus life and, in turn, both student learning and persistence''~\cite{Tinto97-CAC}. In other words, when dealing with networks of peer-to-peer interactions, it is important to consider the environment that fosters their evolution. The in-class networks emerge from individuals who are located in the same time and space (i.e., the classroom during class time) and may even be encouraged to work together (e.g., through group activities). In such a conducive environment where face-to-face interactions take place on a regular basis, one expects a robust network to develop. Such networks can be thought of as ``networks of convenience''. 

Out-of-class networks, on the other hand, are made up of individuals who typically choose to work together. They have to put much more effort into organizing and managing their interactions, be it in person or through some other media (e.g., messengers, text messages, on-line discussion boards, or web-based office suites that allow for collaborating with others in real time). Moreover, the out-of-class network can include not only students enrolled in the course, but also peers who took similar course in the past, friends enrolled in a different section of the same course, faculty that a student is comfortable interacting with, or family members. Thus, the out-of-class networks should be thought of as ``networks of choice''.

\begin{figure}[t]
\centering
\includegraphics[width=.8\linewidth]{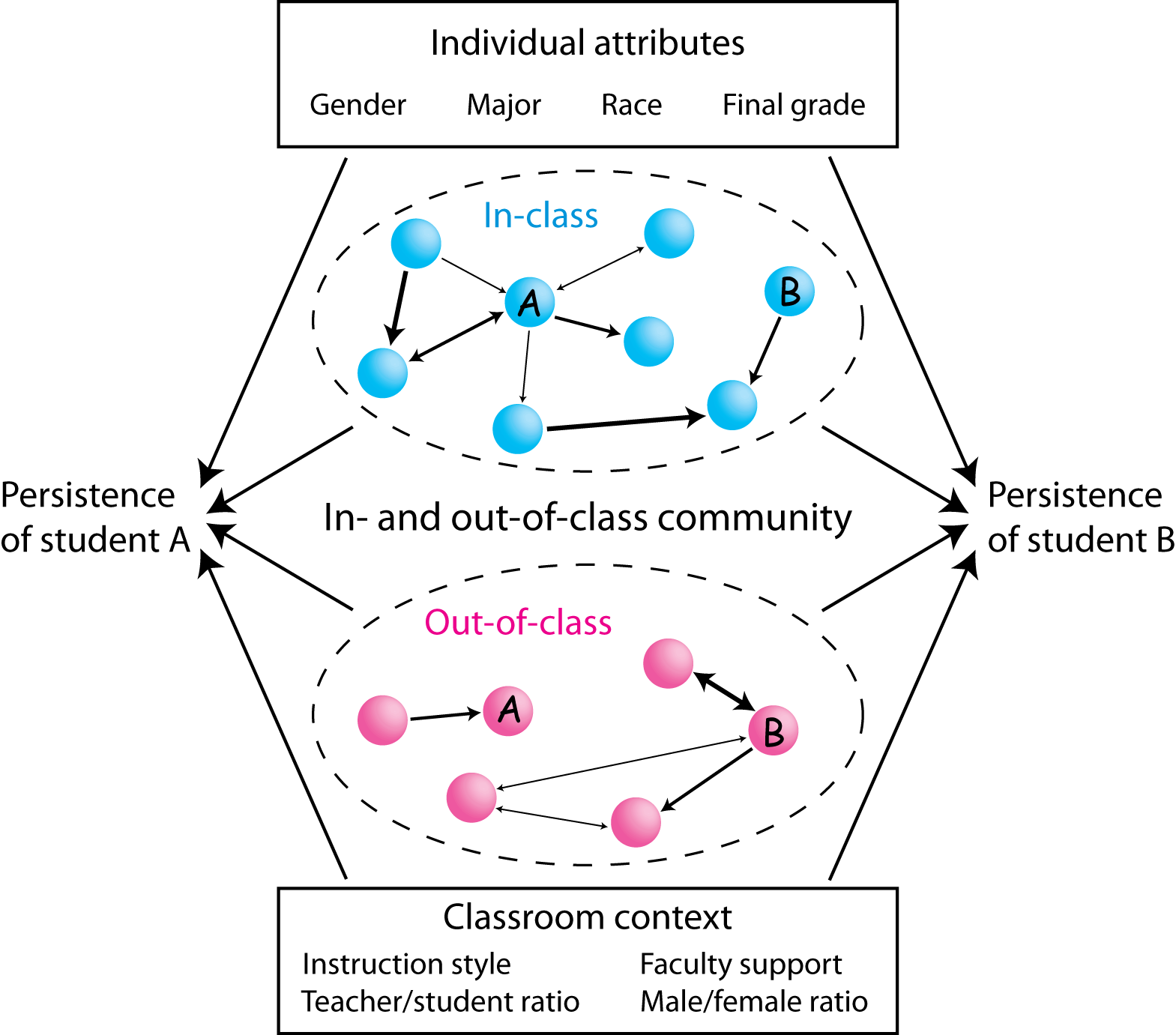}
\caption{Model of persistence. We expand the model of persistence proposed in Ref.~\cite{Zwolak17-IIP} to include frequency of the interactions and out-of-class networks. The categories of factors that may affect students' persistence are individual attributes, classroom context, as well as in- and out-of-class community.}
\label{fig:persist_model}
\end{figure}

In their analysis, Zwolak {\it et al.} use students' embeddedness within networks that include only the presence or absence and the direction of interactions, but not their frequency (i.e., they consider directed binary networks)~\cite{Zwolak17-IIP}. We expand their model of persistence to more fully capture the multilevel character of the network data and to incorporate the out-of-class interactions; see Fig.~\ref{fig:persist_model}. It has been noted that patterns of relations among students might explain outcomes, such as persistence or performance, over and above the attributes of either the individuals or the environment~\cite{Wasserman94}. Our goal is to build an understanding of how students' proximity to others within the in- and out-of-class networks affects---or at least correlates with---their decision about whether to continue in the introductory physics course sequence. 

Moreover, prior studies found that the lack of an engaging pedagogy that promotes active participation and the highly competitive atmosphere in introductory science courses tend to discourage students from pursuing STEM majors~\cite{Gainen95-BTS}. Over the past two decades, active learning strategies have become more common in the classroom~\cite{Freeman14-AEP,Eberlein08-PES,Armbruster09-ALB,Smith05-PEC,Manduca17-IUE}. The number, and presumably quality, of interactions between peers is much higher in these settings than in traditional, lecture-based courses (see, e.g., Table IV in Ref.~\cite{Zwolak17-IIP} and Table I in Ref.~\cite{Brewe10-PLC}). The relative effect of in- and out-of-class interactions, though, has yet to be addressed. 

We focus on the issue of students' persistence within an introductory Modeling Instruction (MI) physics sequence. MI is an approach that strongly emphasizes active learning and can be employed in any subject. The course is interaction driven at both the small and large group levels. There are many ways for students to be involved and make connections---from group activities to group lab reports and, in this course, one group exam to the so-called ``board meetings''~\cite{Brewe08-MTA}. Because of the interactive nature of the course, one expects the in-class network to develop quickly and to include all students enrolled in the course, as well as the instructional staff. Moreover, the nurturing collaborative environment of the MI classroom should promote the culture of working with peers also outside of class. Thus, the MI classroom is an important setting for determining the relevance of the networks of convenience and networks of choice on persistence.

\section{Materials and Methods}\label{sec:methods}
\subsection{The SNA survey}
We collected social network data using a pencil and paper survey developed for this purpose. The survey was given five times throughout the semester. Every 2--3 weeks students were asked two questions regarding in- and out-of-class interactions with their peers. For the in-class part, students were presented with an alphabetized by first name list of names of all classmates, as well as the instructional staff. To simplify the survey completion, they could use two-digit codes randomly assigned to each person instead of full names. To capture the significance of students' in-class interactions, we presented them with a table where, based on repeatability of a given interaction, they were supposed to write names (or codes) of their peers in relevant columns (see Appendix S1 in the Supplemental Material~\cite{supp-mat} for a template of the social network survey). Below the question about out-of-class networks was a blank space where students could provide the relevant information.

\subsection{Survey population}
Data collection for this study spanned two semesters (Fall 2015 and Fall 2016) and four sections of the Modeling Instruction Mechanics (MI-M) at Florida International University. The courses ranged in size from 53 to 74 students. There were 273 students enrolled in MI-M course (147 in Fall 2015, 126 in Fall 2016). Of those, 212 students took a second semester of Modeling Instruction Electricity and Magnetism (MI-EM) course (108 in Spring 2016, 104 in Spring 2017). Table~\ref{tab:survey-pop} includes details about the students' enrollment and teaching staff information for each section. There were three instructors teaching the course (denoted in Table~\ref{tab:survey-pop} as A, B, and C), all physics education researchers. The instructors were accompanied by learning assistants (LAs) and teaching assistants (TAs), i.e., graduate students in physics education and high-achieving undergraduate students who took the MI course before, respectively. The TAs and LAs, as well as their numbers, varied between sections.

\begin{table}[t]
\renewcommand{\arraystretch}{1.2}
\renewcommand{\tabcolsep}{10pt}
\caption{Student enrollment and teaching staff for the MI-M courses. There were two sections of MI-M in Fall 2015 and two in Fall 2016. There were three instructors teaching the course (denoted by A, B, and C). The LAs and TAs varied between semesters.}
\centering
\begin{tabular}{p{2.8cm}cccc} \hline \hline
& \multicolumn{2}{c}{Fall 2015} & \multicolumn{2}{c}{Fall 2016} \\ \hline
Instructor & A & B & A & C \\ 
Number of students & 73 & 74 & 73 & 53 \\
Number of TAs & 1 & 2 & 1 & 0 \\  
Number of LAs & 3 & 3 & 2 & 3 \\ 
 \hline \hline
\end{tabular}
\label{tab:survey-pop}
\end{table}

Demographic information, i.e., gender, ethnicity, and academic plan were available for all students (see Table~\ref{tab:survey-ethn} for gender and ethnicity and Table S1 in the Supplemental Material~\cite{supp-mat} for academic plan). Between the four sections there were a total of 8 reported ethnic groups (including ``choose not to report ethnicity''). Since multiple groups had three students or less, instead of considering each ethnicity as a separate category we combined them into two groups: minority and nonminority. Similarly, for academic plan: we had 32 different majors, with 19 options having less than four students. Thus, we avoid excessive fine graining by grouping majors together into 4 distinct categories (basic science, engineering, science other, other, see Table S1 in the Supplemental Material~\cite{supp-mat} for details). We found no significant difference between sections and the population as a whole in terms of gender [chi-square test: $\chi^2(4)=7.42$, $P=0.12$], ethnicity [chi-square test: $\chi^2(4)=5.81$, $P=0.21$] and academic plan [chi-square test: $\chi^2(12)=17.8$, $P=0.12$]. The final grades data were available for all sections ($N=273$) and the midterm scores were available only for two sections from Fall 2015 ($N=144$; two students in section A and one student in section B were missing midterm exam scores). The Kruskal-Wallis rank sum test showed no statistically significant difference in final grades between sections [$\chi^2(3)=4.17$, $P=0.24$] nor in midterm scores between sections in Fall 2015 [$\chi^2(1)=0.44$, $P=0.51$], the sections for which midterm scores were available.

\begin{table}[t]
\renewcommand{\arraystretch}{1.2}
\renewcommand{\tabcolsep}{5pt}
\caption{Students' gender and ethnicity. For each group, the overall number of students is in the first column, the average percentage of students for all sections is in the second column, and the unbiased estimation of standard deviation between section (SD) is in the third column.}
\centering
\begin{tabular}{llcccc} \hline \hline
\multicolumn{2}{l}{Demographics} & $N$ & Mean (\%) & SD (\%) \\ \hline
Gender & Female & 118 & 43.2 & 7.8 \\
& Male   & 155 & 56.8 & 7.8 \\  
&&&&\\ 
Ethnicity & African American & 34 & 12.4 & 1.8 \\
& Asian & 20 & 7.9 & 5.3 \\
& Hispanic & 186 & 67.6 & 6.7 \\
& White & 21 & 7.4 & 3.7 \\
& Other & 12 & 4.8 & 3.4 \\
 \hline \hline
\end{tabular}
\label{tab:survey-ethn}
\end{table}

\subsection{Social network analysis}\label{sub:sna}
Since the interaction data are relational in nature, we use social network analysis to examine students' integration. Centrality is a family of measures that quantify the relative importance of individuals in a network. From a methodological point of view, centralities provide a unique empirical way to understand students' structural integration into social groups, i.e., it sheds light on student integration via their social ties.

Various social relationships bear different meaning. Thus, treating every contact in the same manner may be overly simplistic. To more fully account for the richness of real life we modify the standard measures to include the frequencies (weights) of interactions (ties) between students (nodes) in the MI courses. For directed degrees---measures of social academic popularity in the case of indegree and sociability or influence in the case outdegree---we use the generalization of Opsahl {\it et al.}\ ~\cite{Opsahl10-CWN} that accounts for both the ties' weights (strength) and their number (degree). The proposed measure is a product of the node's directed degree and strength averaged by degree, with the parameter $\alpha$ tuning the relative importance of each factor. Formally,
\begin{equation}
[C_{T}^{\alpha}]^{\nicefrac{\leftarrow}{\rightarrow}}(i)=C_{D}^{\nicefrac{\leftarrow}{\rightarrow}}(i)\cdot\left[\frac{C_{S}^{\nicefrac{\leftarrow}{\rightarrow}}(i)}{C_{D}^{\nicefrac{\leftarrow}{\rightarrow}}(i)}\right]^\alpha,
\end{equation}
where $\alpha\in[0,\infty)$ and $C_{D}^{\nicefrac{\leftarrow}{\rightarrow}}$ is node's directed degree, i.e.,
\begin{equation}
C_{D}^{\leftarrow}(i)=\sum_jx_{ji}\qquad \small[C_{D}^{\rightarrow}(i)=\sum_jx_{ij}\small].
\end{equation}
Here, $x_{ji}$ ($x_{ij}$) take on value 1 if node $j$ sends a tie to node $i$ (node $i$ sends a tie to node $j$), and 0 otherwise. Also, $C_{S}^{\nicefrac{\leftarrow}{\rightarrow}}$ is a directed strength, i.e.,
\begin{equation}
C_{S}^{\leftarrow}(i)=\sum_jw_{ji}\qquad \small[C_{S}^{\rightarrow}(i)=\sum_jw_{ij}\small],
\end{equation}
with $w_{ji}$ ($w_{ij}$) denoting the weight of a tie from node $j$ to node $i$ (from $i$ to $j$), taking value 0 if a tie is not present. It is easy to see that if $\alpha=0$, then $C_{T}^{\alpha} = C_{D}$ and if $\alpha=1$, then $C_{T}^{\alpha} = C_{S}$. When $\alpha\in(0,1)$, having many weak connections is emphasized more than having only a few strong ones (for the same total strength, Eq. (3)). On the other hand, when $\alpha>1$, it is favorable to have a few strong connections (for the same total strength, Eq. (3)).

The weighted extension for closeness---a measure of embeddedness within the entire network and independence---is\vspace{-3mm}
\begin{equation}\vspace{-2mm}
C_{C}^{\alpha}(i)=\Big[\sum_{j=1}^n d_{ij}^{\alpha}\Big]^{-1},
\end{equation}
where $d_{ij}^{\alpha}=\min\left(w_{im}^{-\alpha}+\dots+w_{nj}^{-\alpha}\right)$ is the weighted shortest path linking $i$ and $j$. When $\alpha=0$, the binary version of closeness results (i.e., the weights are ignored), while for $\alpha=1$ only the weights are important. For $\alpha<1$, a shorter path of weak ties is favored over a longer path with strong ties and for $\alpha>1$ the reverse is true (i.e., the number of intermediary nodes is less important than the strength of the ties).

To truly understand one's involvement in a network it is important to consider not only how well connected a given person is, but also how meaningful their connections are. The generalizations above give an approach that captures the nature of interactions of interest. Moreover, these measures allow for control of the relevant impact of both the number of connections, as well as their weight (frequency). It has been argued that sharing of complex knowledge---i.e., knowledge that is hard to articulate or can be acquired only through experience---requires strong ties~\cite{Hansen99-RWT}, while for easily codified or explicit knowledge, weak ties are more important~\cite{Granovetter73-SWT}. Thus, to emphasize the number of ties, while also taking into account their strength, we chose $\alpha=0.5$.

For eigenvector centrality---the measure of influence on a more global scale---we use the standard extension to weighted graphs: for each node, we sum all frequencies of connections to others weighted by their degrees. The eigenvector centrality is given by the solution to
\begin{equation}
\boldsymbol{A}\,\vec{\boldsymbol{C}}_E=\lambda_{max}\,\vec{\boldsymbol{C}}_E,
\end{equation}
where $\boldsymbol{A}$ is the adjacency matrix of the network, with matrix elements $a_{ij}$ taking on the weight value of a tie from node $i$ to node $j$, $\lambda_{max}$ is the greatest eigenvalue of $\boldsymbol{A}$ and $\vec{\boldsymbol{C}}_E$ is an eigenvector of $\boldsymbol{A}$ corresponding to $\lambda_{max}$. The eigenvector centrality of node $i$ is given by the $i^{th}$ component of vector $\vec{\boldsymbol{C}}_E$.

\subsection{Modeling Instruction classroom network}
\begin{figure*}[t]
\centering
\includegraphics[width=17.8cm]{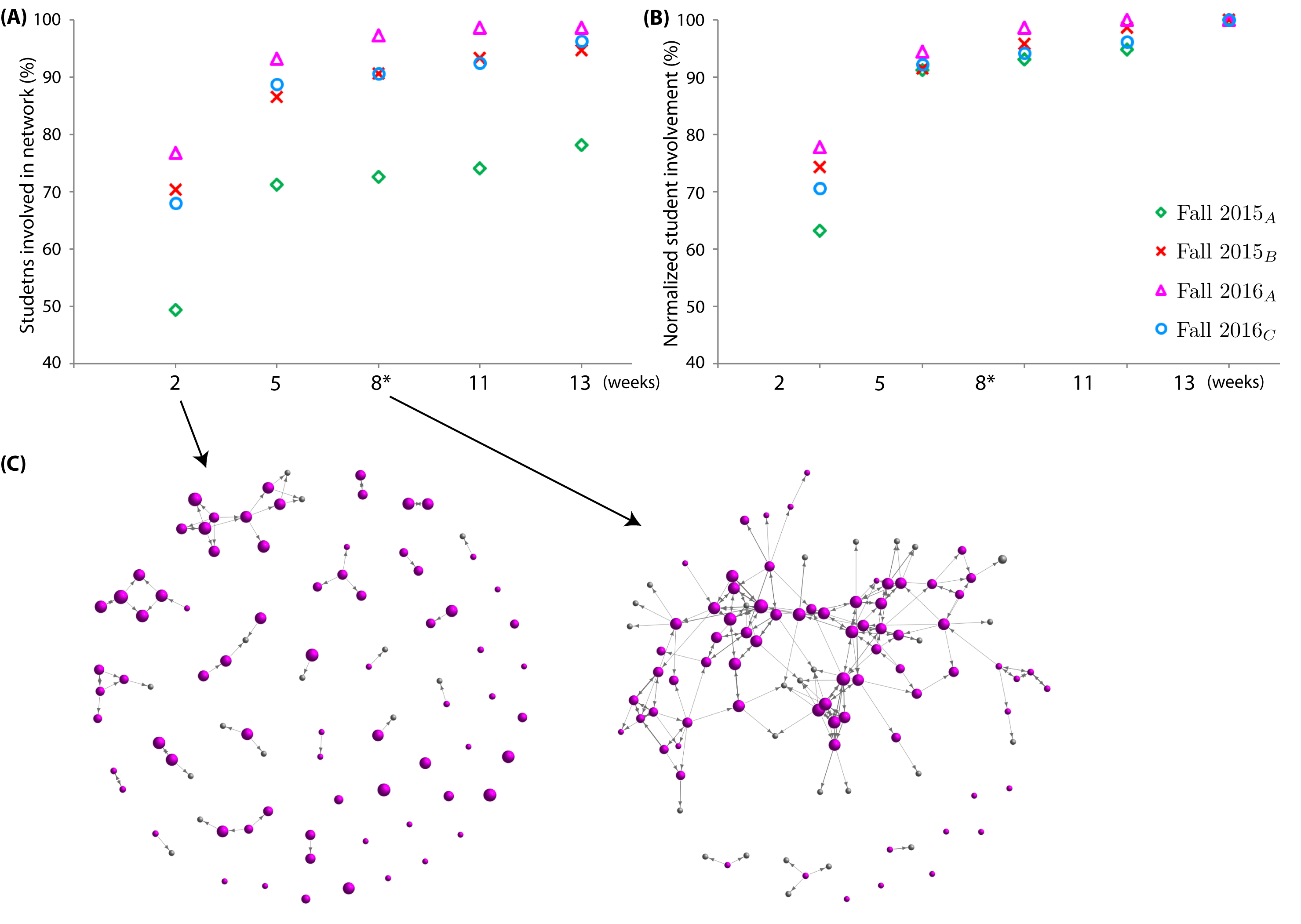}
\caption{Changes in out-of-class network involvement for each section as a function of time. (A) Student involvement in the network versus week (for each of the sections). Student involvement is taken as the percentage of students enrolled in the course who either listed other students or were listed by peers on the out-of-class part of the SNA survey. In three out of four sections, over 90\,\% of students are in the network by week 8 (for the Fall 2015\textsubscript{A} section, the values are consistently lower by about 20\,\%). Note that the third collection was administered in week 8 in three out of the four groups (i.e., Fall 2015\textsubscript{A}, Fall 2015\textsubscript{B}, Fall 2016\textsubscript{A}) and at the beginning of week 9 in one group (i.e., Fall 2016\textsubscript{C}). (B) When normalized by the size of the network at week 13, it becomes clear that the network evolution is comparable between sections and that the equilibrium is basically reached by mid-semester (week 8, the third collection). (C) An example of the out of class network at week 2 and 8 for section Fall 2015\textsubscript{B} (see Fig. S1 in the Supplemental Material~\cite{supp-mat} for the other sections). At week 2, only a few connections are present. By week 8, however, a strong network of students exists, with only a handful of students not embedded within the network. The size of the nodes in both networks corresponds to closeness at week 8 and the color indicates whether a node is a student (magenta) or a person that was not enrolled in the course (gray).}
\label{fig:net-change}
\end{figure*}

Identifying students who are likely to not persist when there is still time for corrective action gives additional opportunities to retain students. We therefore want to examine data for the earliest possible collection during the semester that gives meaningful information. A recent study found that network data collected as early as week 8 (which is midsemester) leads to informative measures for predicting students' academic performance~\cite{Williams17-EII}. Thus, in our analysis, we focus on the collection that took place at the end of week 8 in three out of the four sections and at the beginning of week 9 in the fourth. Moreover, as can be seen in Fig.~\ref{fig:net-change}, the evolution of the out-of-class networks shows that by midsemester all networks are basically fully developed and nearly all students (about 95\,\%) who will end up being in the network by the end of the semester are already present, further supporting this choice. Since MI-M is one of the first courses students take upon entering college, is very likely that most of them do not know other class members. As a result, the development of the out-of-class network will naturally take some time, as students get to know each other through in-class group work and group assignments and learn the benefits of working together.

In our analysis, we consider four in-class networks and four-out-of-class networks (see Table~\ref{tab:networks-comp} for the basic network characteristics). Centrality measures are calculated separately for each network and then the resulting indices from all sections are aggregated to represent one distinct variable (for each measure). As mentioned above, the frequencies are incorporated in the survey question for the in-class network (see Appendix S1 in the Supplemental Material~\cite{supp-mat}). For the out-of-class network, though, students were asked only to report interactions that took place, but not to rate them. To account for repeated out-of-class interactions we aggregate all ``up-to-date'' networks. We do so by pooling answers to the question about out-of-class interactions from the first (completed at the end of week 2), second (completed at the end of week 5 or at the beginning of week 6), and third (completed at the end of week 8 or at the beginning of week 9) survey. Interactions that took place only once during the first 8 weeks of the semester have weight 1, while reoccurring interactions have higher weights (either 2 or 3). Descriptive statistics for the four centralities are presented in Table~\ref{tab:centralities-comp}. Based on the Shapiro-Wilk test we reject the null hypothesis about the normal distribution for all measures.

\begin{table}[t]
\renewcommand{\arraystretch}{1.2}
\renewcommand{\tabcolsep}{5pt}
\caption{Comparison of network characteristics. Basic network descriptors for the in- and out-of-class networks from the third collection (week 8) for all section: network size ($n$), network density ($\Delta$), diameter (the length of the longest path between two nodes; $D$), average path length (the shortest path between two nodes, averaged over all pairs of nodes; $L$), transitivity (a fraction of nodes that share a neighbor – ``a friend of a friend is also my friend''; Tr), reciprocity (a fraction of reciprocal connections; $\rho^\leftrightarrow$).}
\centering
\begin{tabular}{llcccccc} \hline \hline
\multicolumn{2}{l}{Network} & $n$ & $\Delta$ & $D$ & $L$ & Tr & $\rho^\leftrightarrow$ \\ \hline
In-class & Fall 2015\textsubscript{A} & 78 & 0.070 & 11 & 4.3 & 0.6 & 0.5 \\
& Fall 2015\textsubscript{B} & 80 & 0.087 & 7 & 3.1 & 0.5 & 0.5 \\
& Fall 2016\textsubscript{A} & 77 & 0.081 & 7 & 2.8 & 0.4 & 0.4 \\
& Fall 2016\textsubscript{C} & 57 & 0.118 & 7 & 2.5 & 0.4 & 0.4 \\ 
\\
Out-of-class & Fall 2015\textsubscript{A} & 71 & 0.016 & 6 & 2.3 & 0.3 & 0.4 \\
& Fall 2015\textsubscript{B} & 93 & 0.024 & 14 & 5.0 & 0.4 & 0.5 \\
& Fall 2016\textsubscript{A} & 86 & 0.033 & 9 & 3.7 & 0.2 & 0.3 \\
& Fall 2016\textsubscript{C} & 56 & 0.027 & 9 & 3.4 & 0.2 & 0.4 \\
 \hline \hline
\end{tabular}
\label{tab:networks-comp}
\end{table}

\begin{table}[b]
\renewcommand{\arraystretch}{1.1}
\renewcommand{\tabcolsep}{6pt}
\caption{Descriptive statistics for centralities. Based on the Shapiro-Wilk test (test statistics $W$ and $P$ values, columns 1--2), the null hypothesis about the normal distribution is rejected for all centralities. The median ($M$) and interquartile range (IQR) are used to describe the distribution and dispersion for each measure.}
\centering
\begin{tabular}{p{2.0cm}ccccc} \hline \hline
Centrality  & $W$ & $P$ & $N$ & $M$ & IQR \\ \hline
Indegree  & 0.988 & 0.023 & 273 & 7.75 & 4.14 \\
Outdegree & 0.951 & $<0.001$ & 273 & 10.39 & 10.10 \\
Eigenvector & 0.927 & $<0.001$ & 273 & 0.27 & 0.35 \\
Closeness & 0.832 & $<0.001$ & 273 & 0.44 & 0.14 \\
 \hline \hline
\end{tabular}
\label{tab:centralities-comp}
\end{table}

\subsection{Statistical analysis}
For the statistical analysis of the network data, we use the R programming language~\cite{R}, and, in particular, the igraph~\cite{igraph} and tnet~\cite{tnet} packages. The Shapiro-Wilk test was used to test for normality of the centrality measures, the Kruskal-Wallis rank sum test was used to test for statistically significant differences between sections and between different grade levels, and the chi-squared test was used to test for significant differences between sections and the three grade groups in terms of gender, ethnicity and academic plan. 

The statistical significance of independent variables within nested models was verified using the likelihood ratio test (LRT), with the null hypothesis stating no difference in the fit to the data (and thus the simpler model, with less variables, is better). LRT uses the ratio of the maximized value of the likelihood function for the full model over the maximized value of the likelihood function for the simpler models.

To account for the false discovery rate, the Benjamini-Hochberg procedure was implemented. We consider results with $P<0.05$ as significant. All protocols in the project were approved by the Florida International University Institutional Review Board (IRB-13-0240 exempt, category 2). Details about the statistical analysis procedures are presented below. 

\subsubsection{Logistic regression analyses} 
The dependent variable in our study is dichotomous (persistence or lack thereof). To measure the relationship between students' centralities, background information and persistence, we used logistic regression. In the first stage, we wanted to identify centralities that carry significant information about the odds of persisting. To do so, we run 8 simple logistic regression models (four for the in-class and four for the out-of-class centralities) with a single centrality as a predictor. All measures are found to be statistically significant and thus we include all of them in further analysis. Analyses are performed with listwise deletion of missing data. In particular, students who did not report out-of-class interactions on any of the five collections were not included in the out-of-class network and thus centralities for those individuals are not available. The odds ratio (OR) for indegree and outdegree implies how each additional interaction (i.e., increase or decrease of the respective measure by 1, respectively) affects the odds of persisting. The direct effect of individual ties on closeness and eigenvector is much less intuitive and depends strongly on how the topology of the network changes with the additional ties. For example, adding a tie between two otherwise separate groups will have much larger effect on closeness than adding a tie within already a well-connected group~\cite{Magnien15-EIN}. Moreover, both measures are normalized and take on values between 0 and 1, inclusive. Thus, the odds ratio for closeness and eigenvector are rescaled to indicate how odds of persisting change if the measure for an individual increases by 5\,\%. Similarly, OR for grade is rescaled to reflect the increase by a partial letter grade (i.e., from B to B+).

The second stage of our analysis is exploratory. Because of a lack of preexisting theory on how in- and out-of-class centralities should be combined, we consider 16 models with two measures (one of each network type), as well as students' demographic information (i.e., $persistence \sim ic.centrality + oc.centrality + gender + ethnicity + major + grade$). Since we make no assumptions regarding the relationships between the variables, the stepwise regression is an appropriate method for identifying best models. For each model, the Akaike’s Information Criterion (AIC)-based stepwise selection is used to identify all variables that can be removed while minimizing the estimated information loss (implemented in R via the step function). This procedure identifies models that give the best fit to the persistence data without overfitting it (i.e., the removal of the least informative variables). When no more variables can be deleted, the exploratory stage is completed. Our hypothesis is that including an out-of-class centrality in a model should improve its predictive power over only an in-class centrality (and, possibly, demographic information). Thus, we consider for further analysis only models with two centralities and, possibly, other factors.

Comparison of the AIC values allows us to determine the relative loss of information for non-nested models. Based on AIC values for the five models obtained via stepwise-selection we identify three additional models that can be excluded from analysis. The variance inflation factor for the remaining two models, ranging from 1.00 to 1.04, indicates no collinearity between variables.

\subsubsection{Handling missing data}
The network data collections took place in the classroom at the end of class. For the in-class network, the enrollment in the Modeling Instruction course defines the network boundary. That is, all students enrolled are included in the final roster. Since on any given day some of the students where not present and others had to leave the classroom before if officially ended (e.g., to go to their next class), none of the surveys had a $100\,\%$ response rate. The overall response rates, however, were comparable between sections (M = mean, SD = unbiased estimation of a standard deviation: M$_\text{F15A}=84\,\%$, SD$_\text{F15A}=8\,\%$; M$_\text{F15B}=87\,\%$, SD$_\text{F15B}=6\,\%$; M$_\text{F16A}=79\,\%$, SD$_\text{F16A}=11\,\%$; M$_\text{F16C}=80\,\%$, SD$_\text{F16C}=7\,\%$). Using the Kruskal-Wallis rank sum test, we find no significant difference for the response rates between sections [$\chi^2(3)=2.36$, $P=0.50$]. Centralities are fairly robust to random missing data. In particular, for networks with sizes comparable to our data (i.e., 40--75 nodes) up to $35\,\%$ of missing data for directed degrees and about $20\,\%$ for closeness and eigenvector did not affect the overall structure of the network~\cite{Smith13-SES}. Thus, students who do not appear on a given collection are treated as ``isolates'' (disconnected members of a network) and their relevant centrality values are calculated accordingly. For the out-of-class network, though, students who did not appear in the network on any of the five collections are removed from the roster (20 individuals in Fall 2015, 3 individuals in Fall 2016).

\section{Results}
The leading question of our study is as follows: 

\begin{center}
\mbox{\parbox{0.9\linewidth}{\it Do out-of-class networks---the ``networks of choice''---improve models for predicting persistence within the introductory Modeling Instruction sequence?}}
\end{center}

We examine various centrality measures and other factors to determine which ones are the most important. We start by looking at models with a single centrality for the networks (see Sec.~\ref{sub:sna} for explanation of the various centrality measures used in the analysis). Previous studies show that centrality measures tend to be correlated, resulting in redundancy when using more than one in a model~\cite{Bolland88-SOC,Valente08-CCM}. Thus, when considering just one network type (in- or out-of-class), we do not consider models with more than one measure. We  then build more complex models that relay on two centralities (one for in- and one for the out-of-class network), as well as students' information. Our goal is to identify the simplest, accurate model of persistence. 

\subsection{Single predictor models}
In the first stage of our study, we use simple logistic regression models for persistence as predicted by various centralities (i.e., $M_{ic} : persistence \sim ic.centrality$ for in-class models; $M_{oc} : persistence \sim oc.centrality$ for out-of-class models, where $ic.centrality$ and $oc.centrality$ denote measures for the in-class and out-of-class networks, respectively). In particular, we examine indegree, outdegree (local, individual node-level measures), eigenvector (intermediate between local and global measures), and closeness (global measure). Our outcome variable is dichotomous (i.e., persistence or lack of thereof). For both in- and out-of-class networks, those instructors who are reported on the survey become a part of the network. For in-class networks, this results in including all instructional staff (i.e., instructors, TAs and LAs). This is unsurprising since they are an important source of support (be it academic or social) and interactions in class are convenient. For both networks, we want to capture the frequency of interactions, which is done as described in Sec.~\ref{sec:methods}.

Estimates for each measure and their significance levels are in Table~\ref{tab:slrm}. When examining correlations between only persistence and one centrality, all centralities are significant predictors for both in- and out-of-class networks, which is in line with a prior study on in-class networks only~\cite{Zwolak17-IIP}. Thus, we will consider all measures in order to find the most informative. 

\subsection{The power of grade}
In a recent study, Zwolak et al. found that the final grade is the only factor that improves the predictive power of models with an in-class centrality~\cite{Zwolak17-IIP}. A simple logistic regression with grade as a sole predictor for persistence ($M_{gr} : persistence \sim grade$) confirms a positive correlation between final grade and persistence for our data (odd ratio $OR=1.5$; estimate $B=1.36$, standard error of estimate $SE=0.19$, significance level $P<0.001$; sample size $N=273$).

\begin{table}[t]
\renewcommand{\arraystretch}{1.1}
\renewcommand{\tabcolsep}{6pt}
\caption{Logistic regression results for in- and out-of-class centralities. Odds ratio (OR), coefficient estimates and standard error (SE) for the simple logistic regression for persistence as predicted by various in-class and out-of-class centrality measures for the third collection -- the collection in the middle of the term. Significant adjusted $P$ values are marked with an asterisk.}
\centering
\begin{tabular}{p{1.5cm}cccc@{\hskip 5pt}d{3.2}@{\hskip 20pt}c} \hline \hline
\multirow{2}{*}{Centrality} & \multicolumn{3}{c}{Full network} & \multicolumn{3}{c}{Student network} \\ \cline{2-7}
 & OR & B & SE & OR & \multicolumn{1}{c}{B} & SE \\ \hline
Indegree    & 1.21 & $0.19^{***}$ & 0.05 & 1.43 & 0.36^{***} & 0.10 \\ 
Outdegree   & 1.09 & $0.09^{***}$ & 0.02 & 1.26 & 0.23^{**}  & 0.07 \\
Eigenvector & 1.14 & $2.56^{***}$ & 0.75 & 1.16 & 3.00^{*}   & 1.19 \\  
Closeness   & 1.24 & $4.27^{***}$ & 1.03 & 1.38 & 6.48^{***} & 1.48 \\ 
 \hline \hline
\multicolumn{7}{l}{\footnotesize ***$P<0.001$, **$P<0.01$, *$P<0.05$}
\end{tabular}
\label{tab:slrm}
\end{table}

It seems natural that students who score well in class are likely to continue, while students who either fail the class or score poorly will not (and, in cases of failure, they are prohibited from continuing). The behavior of students in the middle of the pack is less obvious. Thus, to gain a better understanding for how final grades affect student's decision whether to persist, we split the data into three groups based on grades. An inspection of final grades leads us to identifying as cutoff points grades B+ and C. This division gives us three fairly equitable groups: ``high'' (grades A, A-, B+), accounting for 34\,\% of grades, ``intermediate'' (grades B, B-, C+), including 37\,\% of grades, and ``low'' (C and lower), including 29\,\% of grades. We found no difference between the groups in terms of gender [chi-squared test: $\chi^2(3)=0.63$, $P=0.89$], ethnicity [chi-squared test: $\chi^2(3)=2.73$, $P=0.43$], and academic plan [chi-squared test: $\chi^2(9)=14.7$, $P=0.10$]. In what follows we focus on the intermediate category, though we will verify applicability of the identified models also for all grade levels.

\begin{table*}[t]
\renewcommand{\arraystretch}{1.2}
\renewcommand{\tabcolsep}{6pt}
\caption{Summary of the likelihood ratio test (LRT) for multiple logistic regression models. Comparison of the full models (with two centralities and final grades as predictors) with the reduced models. The analysis is performed for the intermediate grade level ($N=90$) and for the full data set ($N=239$). dof indicates degrees of freedom. Significant adjusted $P$ values are marked with an asterisk.}
\centering
\begin{tabular}{p{8.5cm}cd{3.2}cd{3.2}cd{3.2}} \hline \hline
\multirow{2}{*}{Full models for persistence} & \multicolumn{2}{c}{$LRT(\text{M}_\text{F}, \text{M}_1)$} & \multicolumn{2}{c}{$LRT(\text{M}_\text{F}, \text{M}_2)$} & \multicolumn{2}{c}{$LRT(\text{M}_\text{F}, \text{M}_3)$}\\ \cline{2-7}
 & dof & \multicolumn{1}{c}{$\chi^2$} & dof & \multicolumn{1}{c}{$\chi^2$} & dof & \multicolumn{1}{c}{$\chi^2$} \\ \hline
\multicolumn{7}{l}{Data set reduced to the intermediate grade level} \\
\hspace{0.25cm}$\text{M}_\text{F}: persistence \sim ic.closeness + oc.outdegree + grade$ & 1 & 2.9 & 1 & 5.9^{*} & 1 & 2.0 \\
\hspace{0.25cm}$\text{M}_\text{F}: persistence \sim ic.outdegree + oc.closeness + grade$  & 1 & 2.7 & 1 & 6.4^{*} & 1 & 2.3  \\ 
&&&&&&\\
\multicolumn{7}{l}{Complete data set} \\
\hspace{0.25cm}$\text{M}_\text{F}: persistence \sim ic.closeness + oc.outdegree + grade$  & 1 & 44.4^{***} & 1 & 10.3^{**} & 1 & 1.8  \\  
\hspace{0.25cm}$\text{M}_\text{F}: persistence \sim ic.outdegree + oc.closeness + grade$  & 1 & 44.6^{***} & 1 & 13.8^{***} & 1 & 1.7 \\ 
 \hline \hline
\multicolumn{7}{l}{\footnotesize ***$P<0.001$, **$P<0.01$, *$P<0.05$}
\end{tabular}
\label{tab:moc-vs-mco}
\end{table*}

\subsection{Analysis of the full models}
Since different centrality measures capture different aspects of networks, and the networks themselves for in- and out-of-class interactions are different in nature (i.e., interactions of convenience versus interactions of choice), it is not obvious how to combine these indices. Thus, in the final step, we examine various combinations of the significant in- and out-of-class centralities. Again, to avoid redundancy, we use only one measure for each network type, i.e., one in-class and one out-of-class, in each model~\cite{Bolland88-SOC,Valente08-CCM}. This gives us a total of 16 models with two measures, as well as students' demographic information (i.e., $M_{F} : persistence \sim ic.centrality + oc.centrality + gender + ethnicity + major + grade$). 

For each model, we employ stepwise-selection based on the AIC to identify all variables that can be removed while minimizing the estimated information loss. This procedure identifies 5 candidate models with three predictors (an in-class centrality, an out-of-class centrality and the final grade) that best fit to the persistence data without overfitting it. Further comparison of the relative quality of these 5 models (as measured by the AIC) allows us to rank them in terms of likelihood of minimizing information loss. In particular, a model that includes in-class outdegree and out-of-class closeness (i.e., $M_{oc} : persistence \sim ic.outdegree + oc.closeness + grade$) is found to give the best fit to the data. The second best model relies on in-class closeness and out-of-class outdegree (i.e., $M_{co} : persistence \sim ic.closeness + oc.outdegree + grade$) and is 0.77 times as probable as the first model to minimize the information loss. The remaining three models perform about 3.2 times worse than the best model and thus we exclude them from further analysis. Consequently, this leaves two candidate models for predicting persistence among students in the middle of the pack. Interestingly, both models include a combination of outdegree and closeness, the only two measures that for the in-class network alone improve the predictive power of grade~\cite{Zwolak17-IIP}.

In the next step, we employ the likelihood ratio test to further investigate the two best models (i.e., $M_{co}$ and $M_{oc}$) for the effect of removing additional variables (one at a time) on the overall fit to the data. We find that, for the intermediate grade cohort, only the out-of-class centrality carries statistically significant information about persistence. Removal of grade or in-class centrality does not affect the fit of either model (see Table~\ref{tab:moc-vs-mco}). The $LRT(M_F, M_1)$, where $M_1 : persistence \sim ic.centrality + oc.centrality$, shows that including information about grade does not improve the model for the intermediate grade level, but it does when all grades are considered. $LRT(M_F, M_2)$, where $M_2 : persistence \sim ic.centrality + grade$, indicates that inclusion of the out-of-class centrality leads to statistically significant improvement of the model fit in both cases. Finally, the $LRT(M_F, M_3)$, with $M_3 : persistence \sim oc.centrality + grade$, implies that the in-class centrality does not improve the predictive power of either of the models. Thus, the outdegree and closeness indices within the network of choice are identified as the most informative measures of persistence for the midrange students. When analyzing all grade levels, removing either the final grade or the out-of-class centrality significantly affects the goodness of the fit, while removal of the in-class centrality does not. In other words, when all grades are considered, the out-of-class closeness or outdegree combined with final grade gives the best prediction of persistence. These are surprising findings that we will discuss momentarily.

\subsection{Model verification}
We determined that, for students with final grades in the intermediate range (i.e., B, B- and C+ grades), the best models of persistence are single-predictor models with either out-of-class outdegree or out-of-class closeness (i.e., $M_{outdeg} : persistence \sim oc.outdegree$ and $M_{close}  : persistence \sim oc.closeness$, respectively). Applying these models to the intermediate final grades confirms that both out-of-class measures are statistically significantly correlated with persistence (see Table~\ref{tab:mod-ver}).

\begin{table}[b]
\renewcommand{\arraystretch}{1.2}
\setlength{\tabcolsep}{6pt}
\caption{Model verification without grade. Odds ratio (OR), estimates (B), and standard error (SE) for simple logistic regression models for persistence as predicted by out-of-class outdegree ($M_{outdeg}$; columns 3--5) and out-of-class closeness ($M_{close}$; columns 6--8). We consider two cases to test our models: (1) the intermediate grades category based on final grades ($N=90$); (2) the intermediate grades category based on midterm scores ($N=47$). Significant $P$ values are marked with an asterisk.}
\centering
\begin{tabular}{p{1.35cm}p{1.24cm}@{\hskip 3pt}c@{\hskip 0pt}d{3.2}c@{\hskip 5pt}c@{\hskip 0pt}d{3.2}@{\hskip 15pt}c} \hline \hline
\multirow{2}{*}{Data} & \hspace{-3mm}\multirow{2}{*}{Coefficient} & \multicolumn{3}{c}{$\text{M}_\text{outdeg}$} & \multicolumn{3}{c}{$\text{M}_\text{close}$} \\ \cline{3-8}
 & & OR & \multicolumn{1}{c}{B} & SE & OR & \multicolumn{1}{c}{B} & SE \\ \hline
Mid-level & \hspace{-3mm}Intercept & & 1.12^{**} & 0.42 & & 0.71 & 0.53 \\ 
(final)   & \hspace{-3mm}Centrality & 1.31 & 0.27^{*} & 0.14 & 1.38 & 6.44^{*} & 2.76\\ \hline
Mid-level & \hspace{-3mm}Intercept &  & 0.73 & 0.50 &  & 0.41 & 0.54 \\ 
(exam) & \hspace{-3mm}Centrality & 1.39 & 0.33 & 0.20 & 1.58 & 9.15^{*} & 4.28 \\ 
\hline \hline
\multicolumn{8}{l}{\footnotesize **$P<0.01$, *$P<0.05$}
\end{tabular} 
\label{tab:mod-ver}
\end{table}

Identifying students who are less likely to persist is important when there is still time to take actions to help them. Thus, using the final grade makes the models ill-suited for practical purposes. Since midterm grades are traditionally used to identify students who are at risk of failing the class (and may also be used to predict persistence), we test our models using midterm scores as a proxy for final grade. While there are three tests throughout the semester in the MI courses, the first midterm occurs around the same time as the third collection. Thus, we use scores from this test to approximate final grades. 

As expected, we find a positive correlation between the midterm scores and final grades (generalized linear model: residual dev. $=104.14$, $dof=1$, $P<0.001$; $N=143$) and midterm scores and persistence (logistic regression: $OR=1.03$; $B=0.10$, $SE=0.02$, $P<0.001$; $N=143$). Moreover, when dividing midterm scores into three categories introduced earlier, we find a positive correlation also between the midterm scores and final grades within the intermediate category (generalized linear model: residual dev. $=8.04$, $d0f=1$, $P<0.001$; $N=56$). Thus, since the midterm scores behave in a similar manner as final grades, they can indeed be used as a proxy for the final grade. Finally, however, there is no statistically significant correlation between midterm scores and persistence within the intermediate group (logistic regression: $OR=1.06$; $B=0.20$, $SE=0.11$, $P=0.06$; $N=56$). The results of the logistic regression for the models with midterm grade in place of final grade are in Table~\ref{tab:mod-ver}. One can see that for the intermediate group based on midterm scores, only closeness remains significant.

When all grades are considered, we found that models with a centrality for out-of-class network and a grade give the best fit to the data (i.e., $M_{og} : persistence \sim oc.outdegree + grade$ and $M_{cg}  : persistence \sim oc.closeness + grade$). We test these two models in a similar manner as we did for the intermediate cohort. We consider two cases: first we use final grades as a predictor, then we take midterm scores as a proxy for final grades. We find that in both cases the centrality and grade are significant predictors for persistence (see Table~\ref{tab:mod-ver-proxy}).

\begin{table}[t]
\renewcommand{\arraystretch}{1.2}
\setlength{\tabcolsep}{6pt}
\caption{Model verification with grade. Odds ratio (OR), estimates (B), and standard error (SE) for simple logistic regression models for persistence as predicted by out-of-class outdegree and grade ($M_{og}$; columns 2--4) and out-of-class closeness and grade ($M_{cg}$; columns 5--7). We consider two test cases for the grade variable: (1) final grades ($N=239$); (2) midterm scores ($N=144$). The odds ratio for grade is rescaled to reflect the increase by a partial letter grade (i.e., from B to B+). Significant $P$ values are marked with an asterisk.}
\centering
\begin{tabular}{p{1.7cm}p{0.5cm}@{\hskip 7pt}d{3.2}@{\hskip 16pt}p{0.6cm}p{0.5cm}@{\hskip 7pt}d{3.2}@{\hskip 16pt}p{0.6cm}} \hline \hline
\multirow{2}{*}{Coefficient} & \multicolumn{3}{c}{$\text{M}_\text{og}$} & \multicolumn{3}{c}{$\text{M}_\text{cg}$} \\ \cline{2-7}
 & OR & \multicolumn{1}{c}{B} & SE & OR & \multicolumn{1}{c}{B} & SE \\ \hline
Intercept     &      & -2.83^{**} & 0.62 &      & -3.40^{***} & 0.70 \\ 
Centrality    & 1.25 & 0.22^{*}   & 0.08 & 1.38 & 6.50^{***}  & 1.77 \\ 
Final grade   & 1.50 & 1.35^{***} & 0.22 & 1.50 & 1.36^{***}  & 0.22 \\ \hline
Intercept     &      & -6.55^{***} & 1.64 &      & -7.22^{***} & 1.76 \\ 
Centrality    & 1.40 & 0.34^{**}   & 0.12 & 1.68 & 10.43^{***} & 3.08 \\ 
Exam grade & 1.04 & 0.12^{***}  & 0.03 & 1.04 & 0.12^{***}  & 0.03 \\ 
\hline \hline
\multicolumn{7}{l}{\footnotesize ***$P<0.001$, **$P<0.01$, *$P<0.05$}
\end{tabular}
\label{tab:mod-ver-proxy}
\end{table}

Finally, to make practical use out of these results, one would like to know, e.g., the value of out-of-class closeness for which students may be at risk of not persisting. When considering grade, it is clear that having a C or less puts one at risk. However, where such a threshold should be drawn (C, C+, B-, etc.), it is not clear since the intermediate grade levels (C+, B-, and B) are not significantly correlated with persistence. For closeness, one can find the threshold value for the intermediate cohort. It comes at about 0.14 [the significance of this split is confirmed by the chi-squared test: $\chi^2(1)=7.46$, $P=0.006$]. Below this value, students have about a 63\,\% chance of persisting. Above this value, students have about a 92\,\% chance of persisting. This large difference is a reflection of the importance of closeness.

\section{Discussion}
Internal communities, such as learning groups within and out of a classroom, are among the most important factors indicating one's integration into the social structure of a university. Previous studies on the effect of integration on persistence found that outdegree and Bonacich's power (a generalization of eigenvector centrality) had a direct, positive effect on students' persistence~\cite{Thomas00-TTB}. This work, though, did not separate the in- and out-of-class networks, a fine graining of the network data that we introduce and find above to be highly informative. Moreover, while these previous findings might guide the development of activities (academic and social) intended to encourage the formation of students' networks ``designed to enhance\dots\,diversity and foster opportunities for nurturing and connecting emerging students social leaders'', the data used in the analysis was gathered once the semester was over, precluding any intervention intended to help at-risk individuals. 

In order to successfully promote persistence at the individual level, be it through structured and purposeful mixing of students in class or handing out group assignments for out-of-class practice, one has to not only identify reliable measures that can indicate students at risk, but also do it in a timely manner, when the semester is still in progress. In our study, we look at the network data at midsemester, when there is still plenty of time to intervene. Additionally, we adjust the network methodology to fully utilize the richness of real interactions, which are weighted in nature.

Although it has been noted that the effect of social and academic integration ``will vary\dots\,inside and outside the classroom''~\cite{Tinto97-CAC}, previous studies did not explicitly distinguish between those two types of interactions~\cite{Thomas00-TTB,Dawson08-RSC}. In an attempt to better understand integration within various types of networks, we explicitly ask students to report both in- and out-of-class interactions. Since current technological advancement diminishes the spatial and temporal limitations for communication, we ask also for reports of the non-face-to-face interactions. Finally, we examine a minority-serving institution, where most students (about 92\,\%) live off-campus~\cite{USnews}, and many are from low income backgrounds (approximately 50\,\% receive Pell grants~\cite{FIU-aid}). For all these reasons, our case expands and, in important ways, complements previous studies that looked at ``upper middle-class, full-time, residential students''~\cite{Thomas00-TTB}. 

The leading question for our analysis was whether inclusion of the out-of-class interactions improves the proposed model of persistence~\cite{Zwolak17-IIP}. To test this hypothesis, we build models with various combinations of in- and out-of-class centralities combined with students' demographic information (i.e., gender, ethnicity, academic plan, final grade). Consistently with previous studies~\cite{Brewe12-PLC}, we find lack of the effect of gender, ethnicity, etc., on the fit to the data. 

Position within the out-of-class ``network of choice'', however, not only improves the model of persistence but is actually more informative than the in-class ``network of convenience'' and, in some cases, grades. In particular, midterm grades usually serve as a proxy for final grades and have traditionally been used to identify students at risk of failing the course. For students performing very well (and therefore likely to persist) and for students who perform rather poorly (and thus are either more likely to drop or---in case of failing grade---cannot continue), the final grade alone is the best indicator of the likelihood of persistence. This remains true when performance throughout the semester is used to approximate final grades. 

For students with average to good performance (i.e., C+, B-, and B)---students that you want to keep but who traditionally are not considered to be at risk of dropping out---the grade alone does not provide sufficient information to determine the odds of persistence. On the contrary, we find no statistically significant correlation between grades ``in the middle of the pack'' and persistence, whether we use final grades or their midterm proxy. Moreover, although in-class centralities were positively correlated with persistence when looked at separately, and were expected to carry the most significance aside from final grade, we found that in comparison with out-of-class centralities they lost significance. Instead, the level of connectedness (as measured by outdegree) and the overall embeddedness within the network (as measured by closeness) for the network of choice are significant predictors for persistence. In other words, students who are most likely to persist are those who by midsemester are well immersed into the out-of-class social system of the university and are successfully reaching out to others. 

More quantitatively, when a student's closeness is below about 0.14, then the chances of persisting are about 63\,\%, but they are about 92\,\% above that value of closeness. Thus, when grades alone do not provide such information, the awareness of a student's out-of-class social and academic integration is the best indicator of the likelihood of persistence. This is especially important for students whose grades place them in the middle of the pack and who do not otherwise seem at risk of not persisting. For those individuals, their position within the network of choice, established as early as midsemester, can be successfully employed to identify students who, despite satisfactory grades, actually need help or encouragement. Providing them with opportunities to build a strong out-of-class network of support may increase their odds of persisting.

One obvious question is why does the in-class network pale in comparison to the out-of-class network? A potential clue comes from the active engagement approach of the MI classroom. When interactions are strongly encouraged---and, indeed, required in some cases---this will tend to ``squeeze'' the number of interactions (ties) into a narrow range: Everyone interacts a lot. Even still, in-class network measures still retain significance on their own. They only lose significance when compared to out-of-class measures. This suggests that it is students who really become part of the social fabric surrounding the university---not just present but developing ties to others---are the ones that become, or are, committed to their education. 

\subsection{Implications}
Building a network of academic or social support is more difficult in commuter and very large schools. Participation in an active-learning course enables students to develop a network of supportive peers that might help them make an easier transition from high school to college and also to integrate them into the community. MI provides an environment where students can learn and build a network at the same time. This is particularly valuable for nonresidential students. 

MI is an example of a curriculum that strongly emphasizes the importance of collaborative learning and, in the process, promotes the culture of working together. From the network of choice evolution, shown in Fig.~\ref{fig:net-change}, one can see that by the third collection---about half way through the semester---the out-of-class networks are basically fully developed. The great majority of students report working together outside of class at least once. 

In large introductory courses, with hundreds of students enrolled in each section, it is much harder to create a collaborative environment~\cite{Brewe10-PLC}. When practicing active-engagement pedagogy is more difficult, promoting out-of-class collaborations can be especially beneficial for students. The significance of outdegree and closeness (but not indegree) for out-of-class models suggests that students' participation in the network of choice provides them not only with the academic support, but also the support needed to balance the struggles faced in class (or in personal life). In turn, access to such support might influence students' decision to continue in college, despite struggles and challenges. Interviews with students further support this hypothesis~\cite{Williams17-PPS}.

Our findings shed light on one of pressing questions of our time: How to increase persistence in STEM education? Modern technologies allow for fast electronic collection and analysis of network data. Knowing when is the best time for data gathering allows one to conduct the collection only once. Network surveys can be taken by students at the end of a lecture, on their phones or laptops. Then, the data can be processed electronically to obtain the metrics of interest, placing no additional burden on instructors. Regardless of whether a passive (e.g., designing the course or coursework) or active (e.g., encouraging individual students) approach is taken, the results presented above suggest that promoting of out-of-class interactions is a simple method to increase persistence (and, ultimately, the number of STEM degrees awarded).

\subsection{Limitations and next steps}
Although the data for our analysis were collected in an introductory physics course, students enrolled in the course spanned over 20 different majors across the university. Thus, they constitute a representative sample of students enrolled in STEM majors. With that being said, the clear next step is to extend this approach to different areas of study and to more traditional curricula. This will help to determine the utility of SNA for increasing and understanding students' persistence more broadly. Importantly, determining whether the network of choice (opposed to the network of convenience) is the most significant factor will be helpful in employing these findings to increase persistence. 

In addition to major and pedagogical approach, institutional properties may also be important. We observed that the development of the out-of-class network takes about 8 weeks in the semester system. Thus, it is important to examine different academic calendar systems, such as quarters and trimesters, as there is less time for the network to fully evolve. As well, since FIU is a Hispanic-Serving Institutions, our study includes mostly Hispanic and Latina/o student population. While this complements previous studies on predominantly nonminority populations, all groups should be considered in future work.

\begin{acknowledgments} 
We would like to thank faculty members for facilitating data collection. The research was funded by the NSF under the Division of Physics Grant No. 1344247.
\end{acknowledgments}

\newpage


\section*{Supplementary material}

**Supplemental Material**

Appendix S1. The social network survey.

Figure S1. Comparison of the out of class network at week 2 and 8 for all sections.

Table S1. Students' declared majors.

**Appendix S1. Social Network Analysis Survey.** The survey was designed to collect the data about students' interpersonal interactions about physics coursework inside and outside of class.

Student ID:_____          Name:_____

## Networks Survey – Physics

*Thank you for taking this survey. By completing it you contribute to research on the impact of social interactions in physics courses. We are interested in how networks form in and outside of classes. Please answer honestly and to the best of your knowledge.*

*Your answers will be kept anonymous and will not affect your success in this course or your classmates' success. Also, please note that students that you list will not know that you listed them in this survey and you will not know if anyone listed you.*

*For your convenience, we provided the names of all your classmates and instructors below. If you are not exactly sure of a name, choose your best guess. If you don't see a name on the list, you can still write it in the table.*

***Make sure to put your name and Panther ID on both pages!!!***

1/3

16. First STUDENT
11. Second STUDENT
37. Third STUDENT
09. Fourth STUDENT
42. Fifth STUDENT
01. Sixth STUDENT
17. Seventh STUDENT

-- Instructors --
85. First.name LAST-NAME (PROF)
86. First.name LAST-NAME (TA)
87. First.name LAST-NAME (LA)
88. First.name LAST-NAME (LA)

2/3

Student ID:_____          Name:_____

<u>Question 1:</u> Please choose from the presented list people from your physics class that you had a meaningful interaction with **in class** this week, even if you were not the main person speaking or contributing. You may include names of students outside of the group you usually work with. You don't have to fill in all columns. You may use the name of a student or their corresponding number.

| I had a meaningful interaction with these people this week <u>in class</u>... | | |
| --- | --- | --- |
| … one time. | … more than one time but NOT every day. | … every day. |
| | | |

<u>Question 2:</u> Throughout this week, did you work with anyone on physics-related material **outside of class**, either in person or virtually (using, e.g., WhatsApp, Google Chat, etc.)? If so, whom did you work with? (*Please provide their FIRST and LAST name. If you worked with someone not from your class, provide their major if possible.*) If not, please write NONE. If you worked with someone from a school-related group you participate in, please name that group. Also, please name instructors or TAs if you met with one or more out-of-class.

3/3

**Fig. S1. Comparison of the out of class network at week 2 and 8 for all sections.** Student involvement in the out of class network at week 2 and 8 for Fall 2015A **(A)**, Fall 2016A **(B)** and Fall 2016C **(C)**. In each case, the number of connections at week 2 (left) is much lower than at week 8 (right). The size of the nodes in both networks corresponds to closeness at week 8 and the color indicates whether a node is a student (magenta) or a person that was not enrolled in the course (grey).

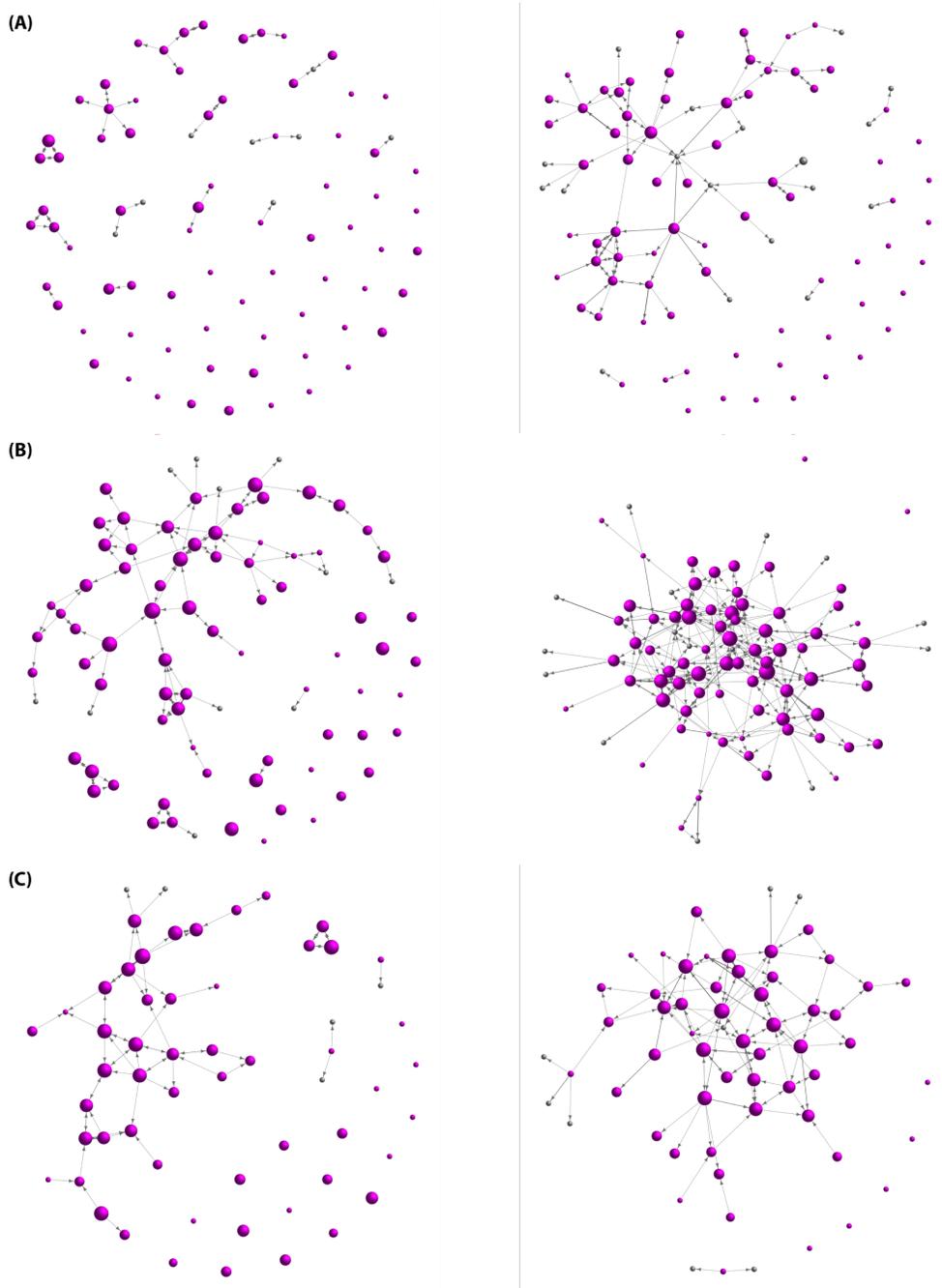

**Table S1. Students' declared majors.** The categories of major are in the first column and the majors in each category are in the second column. For each original major, the overall number of students is in the third column, the average percentage of students for all sections is in the fourth column, and the unbiased estimation of standard deviation between section (SD) is in the fifth column.

| Major-CODE | Major | N | Mean (%) | SD (%) |
|---|---|---|---|---|
| SCIENCE-BASIC | BIOL:BS | 57 | 21.3 | 6.7 |
| | CHEM:BA | 5 | 1.7 | 1.7 |
| | CHEM:BS | 5 | 1.8 | 0.7 |
| | COMPSC:BS | 28 | 10.5 | 3.5 |
| | MATH:BA | 1 | 0.3 | 0.7 |
| | MATH:BS | 1 | 0.3 | 0.7 |
| | PHY:BA | 2 | 0.7 | 0.8 |
| | PHY:BS | 1 | 0.3 | 0.7 |
| ENGENEERING | BIOMEG:BS | 11 | 4.0 | 1.1 |
| | CIVLEG:BS | 12 | 4.2 | 3.2 |
| | COMPEG:BS | 10 | 3.5 | 5.0 |
| | ELEG:BS | 5 | 1.8 | 1.7 |
| | ENVEG:BS | 3 | 1.0 | 0.7 |
| | MECHEG:BS | 24 | 8.6 | 2.7 |
| SCIENCE-OTHER | BIOCHM:BS | 1 | 0.5 | 0.9 |
| | EXP-BIOENV | 1 | 0.3 | 0.7 |
| | GEOSC:BS | 2 | 0.7 | 0.8 |
| | IT:BS | 2 | 0.7 | 0.8 |
| | MRNBIO:BS | 2 | 0.7 | 0.8 |
| OTHER | ACCT:BACC | 1 | 0.3 | 0.7 |
| | DUALAAA | 1 | 0.5 | 0.9 |
| | DUALFIU | 13 | 4.7 | 1.4 |
| | DUALFL | 2 | 0.8 | 1.0 |
| | DUALHS | 51 | 18.3 | 6.1 |
| | EXP-NURSHS | 1 | 0.3 | 0.7 |
| | EXP-PHSCEN | 16 | 6.8 | 8.6 |
| | HSA:BHSA | 1 | 0.3 | 0.7 |
| | NDUGSTU | 1 | 0.3 | 0.7 |
| | PSYC:BA | 9 | 3.1 | 2.3 |
| | SPEC | 1 | 0.5 | 0.9 |
| | SPEC20DUAL | 1 | 0.3 | 0.7 |
| | TRANSIENT | 2 | 0.7 | 0.8 |



\begin{thebibliography}{13}
\expandafter\ifx\csname natexlab\endcsname\relax\def\natexlab#1{#1}\fi
\providecommand{\enquote}[1]{``#1''}
\expandafter\ifx\csname url\endcsname\relax
  \def\url#1{\texttt{#1}}\fi
\expandafter\ifx\csname urlprefix\endcsname\relax\def\urlprefix{URL }\fi
\providecommand{\eprint}[2][]{\url{#2}}

\bibitem{Adkins12-ANS}
Adkins RC (2012) America Desperately Needs More STEM Students. Here's How to Get Them. \emph{Forbes Leadership Forum}. Available at \href{www.forbes.com/sites/forbesleadershipforum/2012/07/09/america-desperately-needs-more-stem-students-heres-how-to-get-them/#5b20132a7497}{www.forbes.com/sites/forbesleadershipforum/2012/07/09 /america-desperately-needs-more-stem-students-heres-how-to-get-them/\#5b20132a7497} (accessed: May 8, 2017).

\bibitem{PCAST12-ETE}
President's Council of Advisors on Science and Technology, Engage to excel: Producing one million additional college graduates with degrees in science, technology, engineering, and mathematics (Executive Office of the President, 2012). Available at \href{eric.ed.gov/?id=ED541511}{eric.ed.gov/?id=ED541511} (accessed: May 8, 2017).

\bibitem{NSF96-STF}
National Science Foundation (NSF), Shaping the future: New expectations for undergraduate education in science, mathematics, engineering, and technology (nsf96139, NSF, 1996). Available at \href{www.nsf.gov/pubs/stis1996/nsf96139/nsf96139.txt}{www.nsf.gov/pubs/stis1996/nsf96139/nsf96139.txt} (accessed: May 8, 2017). 

\bibitem{Tinto75-DHE}
Tinto V (1975) Dropout from higher education: A theoretical synthesis of recent research. {\it Rev. Educ. Res.} {\bf 45}, 89--125. 

\bibitem{Thomas00-TTB}
Thomas SL (2000) Ties that bind: A social network approach to understanding student integration and persistence. {\it J. High. Educ.} {\bf 71}, 591--615. 

\bibitem{Dawson08-RSC}
Dawson S (2008) A study of the relationship between student social networks and sense of community. {\it J. Educ. Techno. Soc.} {\bf 11}, 224--238.

\bibitem{Zwolak17-IIP}
Zwolak JP, Dou R, Williams EA, Brewe E (2017) Students' network integration as a predictor of persistence in introductory physics courses. {\it Phys. Rev. Phys. Educ. Res.} {\bf 13}, 010113. 

\bibitem{Swail04-ASR}
Swail WS (2004) The art of student retention: A handbook for practitioners and administrators (Educational Policy Institute, Austin, TX).

\bibitem{Baldwin97-NES}
Baldwin TT, Bedell MD, Johnson JL (1997) The social fabric of a team-based M.B.A. program: Network effects on student satisfaction and performance. {\it Acad. Manag.~J.} {\bf 40}, 1369--1397. 

\bibitem{Tinto97-CAC}
Tinto V (1997), Classrooms as communities: Exploring the educational character of student persistence. {\it J. High. Educ.} {\bf 68}, 599--623.

\bibitem{Wasserman94}
Wasserman S, Faust K (1994) {\it Social Network Analysis: Methods and Applications} (Cambridge Univ. Press, Cambridge). 

\bibitem{Gainen95-BTS}
Gainen J (1995) Barriers to success in quantitative gatekeeper courses. {\it New Directions for Teaching and Learning} {\bf 1995}, 5--14.

\bibitem{Freeman14-AEP}
Freeman S, Eddy SL, McDonough M, Smith MK, Okoroafor N, Jordt H, Wenderoth MP (2014) Active learning increases student performance in science, engineering, and mathematics. {\it Proc. Natl. Acad. Sci. U.S.A.} {\bf 111}, 8410--8415.

\bibitem{Eberlein08-PES}
Eberlein T, Kampmeier J, Minderhout V, Moog RS, Platt T, Varma-Nelson P, White HB (2008) Pedagogies of engagement in science: A comparison of PBL, POGIL, and PLTL. {\it Biochem. Mol. Biol. Educ.} {\bf 36}, 262--273.

\bibitem{Armbruster09-ALB}
Armbruster P, Patel M, Johnson E, Weiss M (2009) Active Learning and Student-centered Pedagogy Improve Student Attitudes and Performance in Introductory Biology. {\it CBE Life Sci. Educ.} {\bf 8}, 203--213.

\bibitem{Smith05-PEC}
Smith KA, Sheppard SD, Johnson DW, Johnson RT (2005) Pedagogies of Engagement: Classroom-Based Practices. {\it J. Eng. Educ.} {\bf 94}, 87--101.

\bibitem{Manduca17-IUE}
Manduca CA, Iverson ER, Luxenberg M, Macdonald RH, McConnell DA, Mogk DW, Tewksbury BJ (2017) Improving undergraduate STEM education: The efficacy of discipline-based professional development. {\it Sci. Adv.} {\bf 3}, e1600193. 

\bibitem{Brewe10-PLC}
Brewe E, Kramer LH, O'Brien GE (2010) in {\it AIP Conference Proceedings} {\bf 1289}, 85--88. 

\bibitem{Brewe08-MTA}
Brewe E (2008) Modeling theory applied: Modeling instruction in introductory physics. {\it Am. J. Phys} {\bf 76}, 1155--1160.

\bibitem{supp-mat}
See Supplemental Material at \href{http://link.aps.org/ supplemental/10.1103/PhysRevPhysEducRes.14.010131}{http://link.aps.org/ supplemental/10.1103/PhysRevPhysEducRes.14.010131} for the social network survey, comparison of the out-of-class networks for all sections and for students’
declared major grouping.

\bibitem{Opsahl10-CWN}
Opsahl T, Agneessens F, Skvoretz J (2010) Node centrality in weighted networks: Generalizing degree and shortest paths. {\it Soc. Networks} {\bf 32}, 245--251.

\bibitem{Hansen99-RWT}
Hansen MT (1999) The search-transfer problem: The role of weak ties in sharing knowledge across organization subunits. {\it Admin. Sci. Q.} {\bf 44}, 82--111.

\bibitem{Granovetter73-SWT}
Granovetter MS (1973) The Strength of Weak Ties. {\it Am. J. Sociol.} {\bf 78}, 1360--1380.

\bibitem{Williams17-EII}
Williams EA, Zwolak JP, Dou R, Brewe E (2017), Student engagement as a predictor of academic performance in the introductory physics classroom. arXiv:1706.04121.

\bibitem{Smith13-SES}
Smith JA, Moody J (2013) Structural effects of network sampling coverage I: Nodes missing at random. {\it Soc. Networks} {\bf 35}, 652--668.

\bibitem{R}
R Core Team (2016) R: A Language and Environment for Statistical Computing. R Foundation for Statistical Computing, Vienna, Austria.

\bibitem{igraph}
Csardi G, Nepusz T (2006) The igraph software package for complex network research. InterJournal, Complex Systems 1695.

\bibitem{tnet}
Opsahl T (2009) Structure and Evolution of Weighted Networks. University of London (Queen Mary College), London, UK.

\bibitem{Magnien15-EIN}
Magnien C, Tarissan F (2015) Time evolution of the importance of nodes in dynamic networks. [in:] {\it Proceedings of the 2015 IEEE/ACM international conference on Advances in Social Networks Analysis and Mining (ASONAM)}, 1200--1207.

\bibitem{Bolland88-SOC}
Bolland JM (1988) Sorting Out Centrality: An analysis of the performance of four centrality models in real and simulated networks. {\it Soc. Networks} {\bf 10}, 233--253.

\bibitem{Valente08-CCM}
Valente TW, Coronges K, Lakon C, Costenbader E (2008) How Correlated Are Network Centrality Measures? {\it Connections} {\bf 28}, 16--26. 

\bibitem{USnews}
US News: America's best colleges. Available at \href{http://colleges.usnews.rankingsandreviews.com/best-colleges/fiu-9635}{http://colleges.usnews.rankingsandreviews.com/best-colleges/fiu-9635} (accessed: October 12, 2016).

\bibitem{FIU-aid}
Office of Governmental Relations, Florida International University. Available at \href{http://government.fiu.edu/federal/financial-aid/index.html}{http://government.fiu.edu/federal/financial-aid/index.html} (accessed: May 8, 2017).

\bibitem{Brewe12-PLC}
Brewe E, Kramer LH, Sawtelle V (2012) Investigating student communities with network analysis of interactions in a physics learning center. {\it Phys. Rev. ST Phys. Educ. Res.} {\bf 8}, 010101.

\bibitem{Williams17-PPS}
Williams EA, Zwolak JP, Brewe E (2017) Physics major engagement and persistence: A phenomenography interview study. [in:] {\it Proceedings of the Physics Education Research Conference 2017}, Cincinnati, OH (AIP, New York, 2018), 436--439.
\end{thebibliography}
\end{document}